\documentclass[usenatbib,useAMS]{mn2e}

\usepackage{epsf}
\usepackage[dvips]{epsfig}
\usepackage{times}

\def\spose#1{\hbox to 0pt{#1\hss}}
\def\lta{\mathrel{\spose{\lower 3pt\hbox{$\sim$}}
    \raise 2.0pt\hbox{$<$}}}
\def\gta{\mathrel{\spose{\lower 3pt\hbox{$\sim$}}
    \raise 2.0pt\hbox{$>$}}}

\begin{document}

\label{firstpage}

\title[The Orphan Stream Progenitor]
{Is Ursa Major II the Progenitor of the Orphan Stream?} 

\author[Fellhauer et al.]
{M. Fellhauer$^{1}$ \thanks{Email:
madf@ast.cam.ac.uk,nwe@ast.cam.ac.uk,vasily@ast.cam.ac.uk},
  N.W. Evans$^{1}$, 
  V. Belokurov$^{1}$, 
  D.B. Zucker$^{1}$,
  B. Yanny$^{2}$,
  \newauthor
  M.I. Wilkinson$^{1}$,
  G. Gilmore$^{1}$,
  M.J. Irwin$^{1}$,
  D.M. Bramich$^{1}$,
  S. Vidrih$^{1}$,
  P. Hewett$^{1}$,
  \newauthor
  T. Beers$^{3}$ \\
  $^{1}$ Institute of Astronomy, University of Cambridge, Madingley
  Road, Cambridge CB3 0HA, UK \\
  $^{2}$ Fermi National Accelerator Laboratory, P.O. Box 500, Batavia,
  IL 60510, USA \\
  $^{3}$ Department of Physics and Astronomy and Joint Institute for
  Nuclear Astrophysics, Michigan State University, East Lansing, MI
  48824, USA}

\pagerange{\pageref{firstpage}--\pageref{lastpage}} \pubyear{2006}

\maketitle

\begin{abstract}
  Prominent in the `Field of Streams' -- the Sloan Digital Sky Survey
  map of substructure in the Galactic halo -- is an `Orphan Stream'
  without obvious progenitor. In this numerical study, we show a
  possible connection between the newly found dwarf satellite Ursa
  Major II (UMa~II) and the Orphan Stream. We provide numerical
  simulations of the disruption of UMa~II that match the observational
  data on the position, distance and morphology of the Orphan Stream.
  We predict the radial velocity of UMa~II as $-100 \ {\rm
  km\,s}^{-1}$, as well as the existence of strong velocity gradients
  along the Orphan Stream.  The velocity dispersion of UMa~II is
  expected to be high, though this can be caused both by a high dark
  matter content or by the presence of unbound stars in a disrupted
  remnant. However, the existence of a gradient in the mean radial
  velocity across UMa~II provides a clear-cut distinction between
  these possibilities.  The simulations support the idea that some of
  the anomalous, young halo globular clusters like Palomar~1 or Arp~2
  or Ruprecht 106 may be physically associated with the Orphan Stream.
\end{abstract}


\section{Introduction}
\label{sec:intro}

Data from the Sloan Digital Sky Survey \citep[SDSS;][]{Yo00} have
revealed abundant examples of streams and substructure in the Milky
Way halo. For example, \citet{Be06a} used a simple colour cut $g-r <
0.4$ to map out the distribution of stars in SDSS Data Release 5
(DR5). The ``Field of Streams'', an RGB-composite image composed of
magnitude slices of the stellar density of these stars, showed the
leading arm of the well-known Sagittarius stream and the Monoceros
ring very clearly. Also prominent was a new stream, which did not have
an identified progenitor, and was called the Orphan Stream
by~\citet{Be06a}.

The Orphan Stream was then analysed independently by two
groups. \citet{Gr06} reported that there was a diminutive Galactic
satellite that lay near the projected path of the new stream but that
it was ``unlikely to be related to it''. \citet{Be06b} disagreed,
noting that the diminutive satellite lay on the same Galactocentric
great circle as the Orphan Stream. They argued that there was a
preponderance of unusual objects along this great circle -- including
the Complex A High Velocity Clouds and the young halo globular
clusters Ruprecht 106 and Palomar 1 -- and suggested that some or all
may be the remnants of the disruption of a dwarf galaxy. Then,
\citet{Zu06} provided follow-up Subaru imaging of the diminutive
satellite, confirming it as a disrupted dwarf galaxy and naming it
Ursa Major II (UMa~II) after its host constellation.

One possible interpretation of the data is that UMa~II is the
progenitor of the Orphan Stream. Closely related is the possibility
that both UMa~II and the Orphan Stream are remnants from the break-up
of a still larger object, perhaps a tidal dwarf galaxy~\citep[see
e.g.,][]{Kr97}. In this theoretical study, we strengthen the case for
such interpretations by providing an orbit for the disruption of
UMa~II so that its tidal tails match the observational data available
on the Orphan Stream.

\section{Observational Data}
\label{sec:obs}

\begin{table}
  \centering
  \caption{Positions, distance moduli, distances and heliocentric
    velocities of the Orphan Stream From Belokurov et al. (2006b).}   
  \label{tab:orphan}
  \begin{tabular}{@{}ccccc}
    $\alpha$ & $\delta$ & $m-M$ & $D_\odot$ & $v_\odot$ \\
    \hline 
    $162.1^\circ$ & $-0.5^\circ$ & $16.5 \pm 0.7$ & $20^{+7}_{-5}$ kpc
    & $-35 \pm 10$ km\,s$^{-1}$ \\ 
    $158.9^\circ$ & $8.5^\circ$  & $16.5 \pm 0.9$ & $20^{+10}_{-7}$
    kpc & --- \\ 
    $155.4^\circ$ & $17.0^\circ$ & $17.1 \pm 0.7$ & $26^{+10}_{-7}$
    kpc & --- \\ 
    $152.3^\circ$ & $25.0^\circ$ & $17.5 \pm 0.8$ & $32^{+13}_{-10}$
    kpc & --- \\ 
    $149.4^\circ$ & $32.0^\circ$ & $17.5 \pm 0.9$ & $32^{+15}_{-12}$
    kpc & $+105 \pm 10$ km\,s$^{-1}$ \\ 
    \hline
  \end{tabular}
\end{table}

\begin{figure*}
  \centering
  \epsfxsize=5.7cm
  \epsfysize=5.7cm
  \epsffile{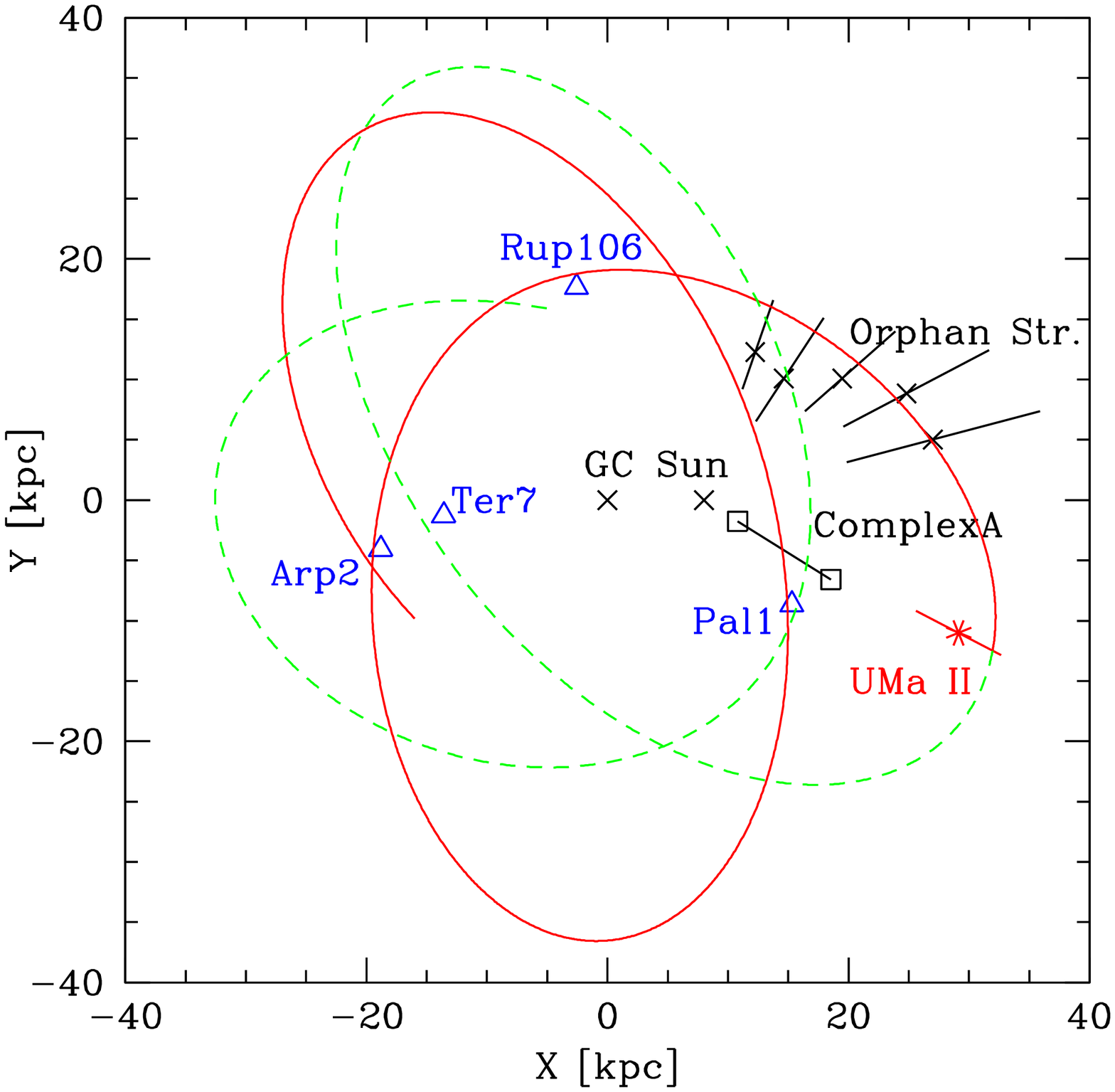}
  \epsfxsize=5.7cm
  \epsfysize=5.7cm
  \epsffile{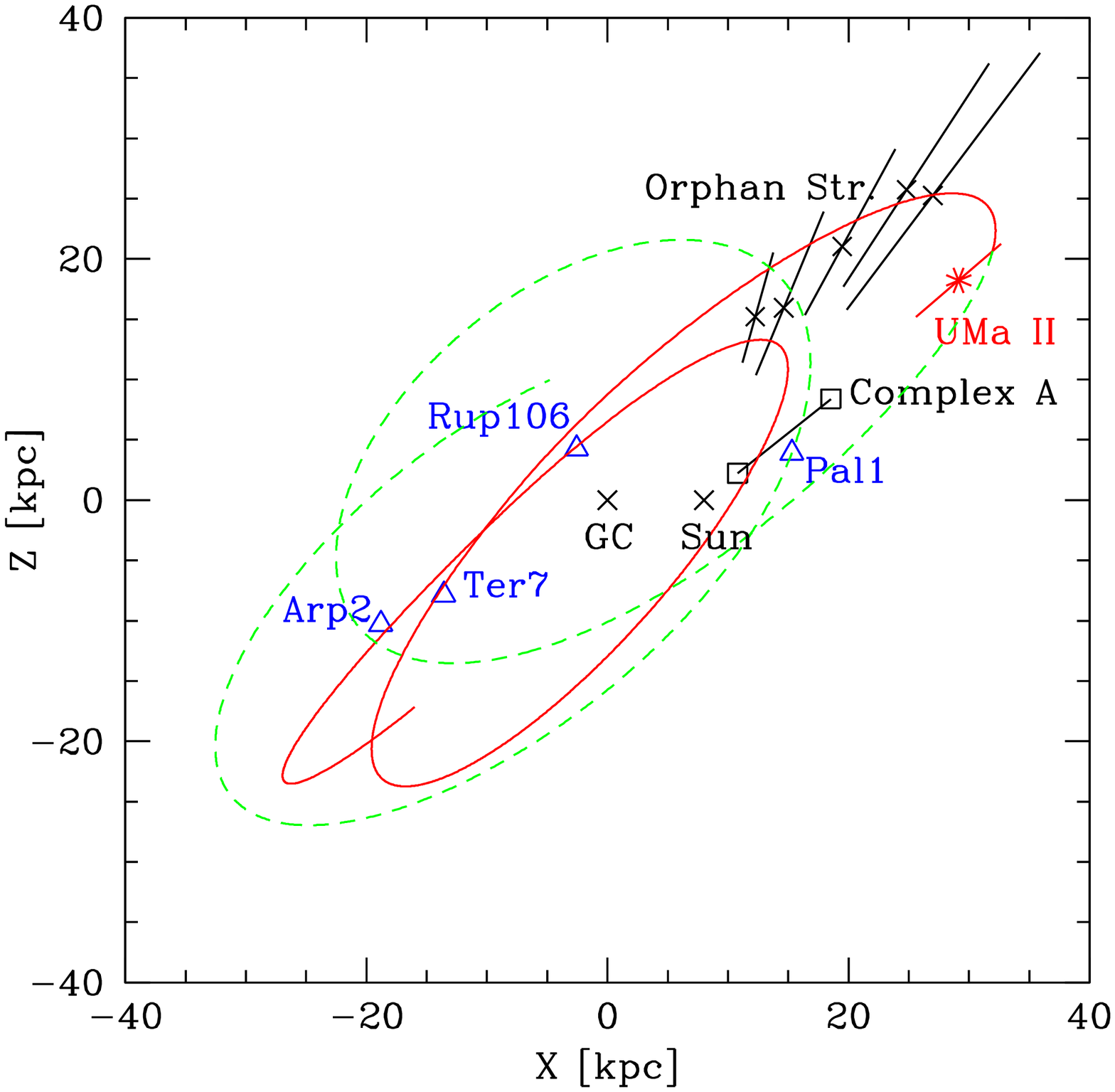}
  \epsfxsize=5.7cm
  \epsfysize=5.7cm
  \epsffile{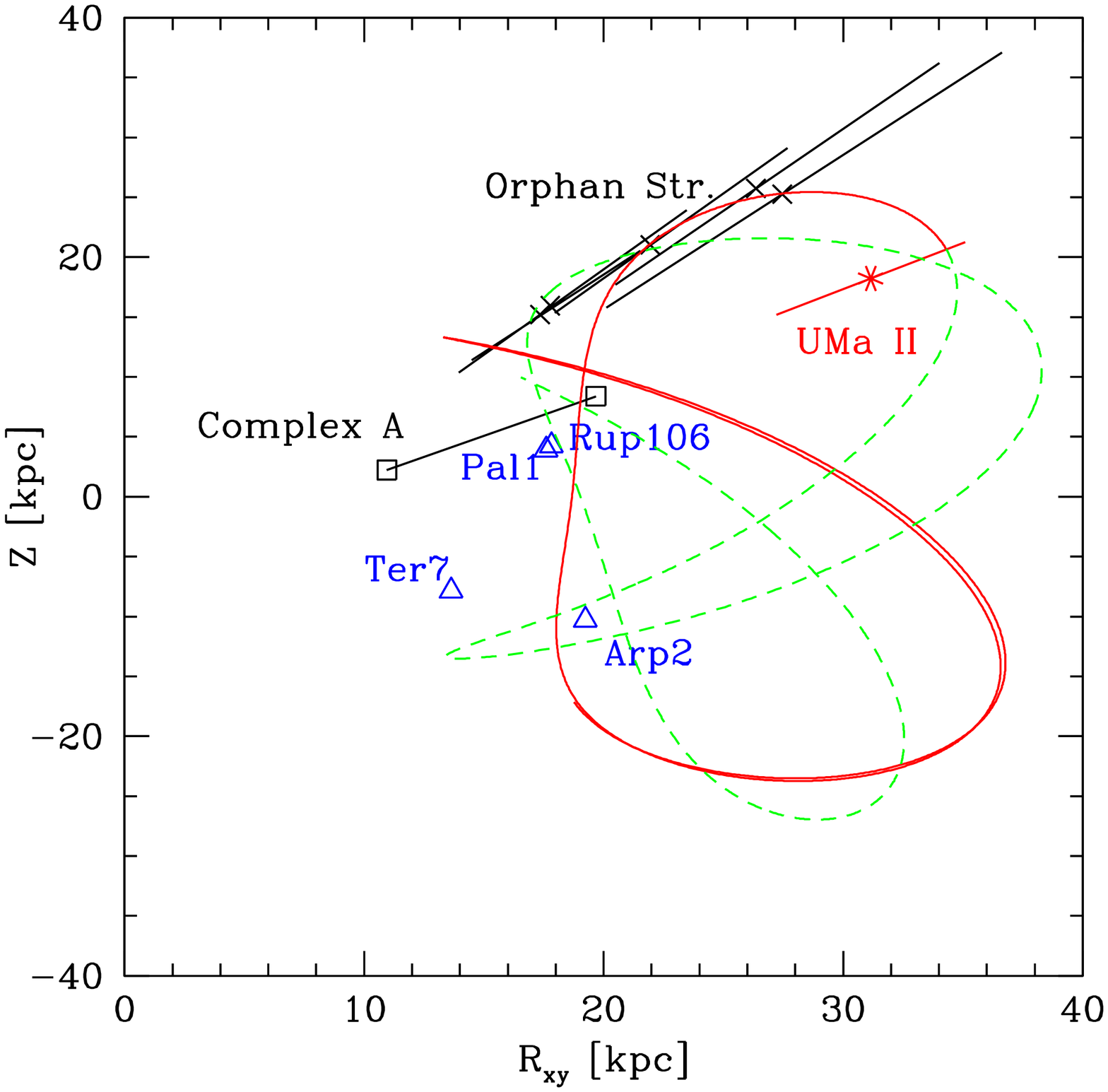}
  \caption{Orbit of UMa~II in the $(x,y)$-plane (left), $(x,z)$-plane
(middle) and $(R= \sqrt{x^{2}+y^{2}},z)$-plane (right). The red solid
line is the backwards orbit from the present position and the green
dashed line is the forward orbit over 1.5 Gyr. The red star with error
bar shows the present position of the UMa~II dwarf galaxy.  Black
crosses with error bars show the position of the Orphan Stream from
Table~\ref{tab:orphan}.  The positions of some globular clusters which
may be associated with the stream are marked with blue, open
triangles.  The distance bracket to the Complex A is marked with
black, open squares.}
\label{fig:orbit}
\end{figure*}

The UMa~II dwarf galaxy \citep{Zu06} is located at right ascension
$\alpha = 132.8^{\circ}$ and declination $\delta = +
63.1^{\circ}$. Its heliocentric distance is estimated as $D_{\odot} =
30 \pm 5 \ {\rm kpc}$, whilst its radial velocity is as yet
unmeasured.  UMa~II appears elongated along lines of increasing right
ascension with an ellipticity of $\sim 0.5$.  Follow-up observations
of the central parts with Subaru reveal more than one density
enhancement within the satellite, which supports the fact that it may
be in the process of tidal disruption.  But, no obvious tails around
the object are discernible in the wider field SDSS data. The total
luminosity of UMa~II is $M_{\rm tot,V} = -3.8 \pm 0.6 \ {\rm mag.}$
This translates into a stellar mass of $\approx 6 \times
10^3$~M$_{\odot}$, applying a conservative mass-to-light ratio of $2$
which is typical for an old population.  This is a lower limit for the
present-day mass of the remnant object.

UMa~II lies on the same great circle as the Orphan Stream.  This can
be traced for over $\sim 50^\circ$ in upper main sequence and turn-off
stars in the SDSS data. By constructing a colour-magnitude mask based
on the ridge-line of the old metal-poor globular cluster M92,
\citet{Be06b} showed that the Orphan Stream is closer to us at lower
declinations than at higher.  The distances and distance moduli to the
Stream at different right ascension and declination are listed in
Table~\ref{tab:orphan}.

The total magnitude of the Orphan Stream is $m_{r} \approx 9.8$.
Assuming the smallest distance modulus from Table~\ref{tab:orphan} of
$m-M = 16.5$, this results in $M_{r} \sim -6.7$ or $3.5 \times 10^{4}$
solar luminosities.  Taking the largest distance modulus of $m-M =
17.5$, the total luminosity of the stream is $\approx 8 \times
10^{4}$~L$_{\odot}$.  With a mass-to-light ratio for an old stellar
population of $2$, this amounts to a total mass in stars in the Orphan
Stream of $\approx 10^{5}$~M$_{\odot}$.

\citet{Be06b} speculated that there might be a connection between the
Orphan Stream and the agglomeration of high velocity clouds known as
Complex A, which lie on the same great circle.  Complex A is located
between $\alpha = 126.7^{\circ}$, $\delta = 67.4^{\circ}$ and $\alpha
= 134.5^{\circ}$, $\delta = 61.7^{\circ}$ with a distance bracket
between $4$ and $15$~kpc in heliocentric distance~\citep[see
e.g.,][]{Wa96,Wa01}.  The measured radial velocity of this cloud
complex is in the range of $-140$ to $-190$~km\,s$^{-1}$.  Of course,
the velocities of gas clouds may be affected by forces other than
gravitational ones.

\section{Set-up}
\label{sec:setup}

Our working hypothesis is that the UMa II dwarf galaxy is the
progenitor of the Orphan Stream.  To determine a possible orbit, we
first perform test-particle integrations in a Milky Way potential
which consists of a logarithmic halo of the form
\begin{eqnarray}
  \label{eq:halopot}
  \Phi_{\rm halo}(r) & = & \frac{v_{0}^{2}}{2} \ln \left( x^{2} + y^2q^{-2}
    + d^{2} \right),
\end{eqnarray}
with $q =1, v_{0}=186$~km\,s$^{-1}$ and $d=12$~kpc.  The disc is represented
by a Miyamoto-Nagai potential:
\begin{eqnarray}  
  \label{eq:discpot}
  \Phi_{\rm disc}(R,z) & = & \frac{G M_{\rm d}} { \sqrt{R^{2} + \left(
        b + \sqrt{z^{2}+c^{2}} \right)^{2}}},
\end{eqnarray}
with $M_{\rm d} = 10^{11}$~M$_{\odot}$, $b = 6.5$~kpc and $c =
0.26$~kpc. Finally, the bulge is modelled as a Hernquist potential
\begin{eqnarray}
  \label{eq:bulgepot}
  \Phi_{\rm bulge}(r) & = & \frac{G M_{\rm b}} {r+a},
\end{eqnarray}
using $M_{\rm b} = 3.4 \times 10^{10}$~M$_{\odot}$ and $a=0.7$~kpc.
The superposition of these components gives quite a good
representation of the Milky Way.  The circular speed at the solar
radius is $\sim 220$~km\,s$^{-1}$.  The major advantage is the
analytical accessibility of all quantities (forces, densities, and so
on).

First, we use trial and error to find a suitable orbit which reproduces
most of the observational data. We then compute this orbit backwards
for $10$~Gyr and insert a live progenitor.  We use the particle-mesh
code Superbox \citep{Fe00} to perform the forward integration until
the position of UMa~II today is reached.  We then analyse the location
of the tidal tails, adjust the parameters from the test-particle
simulation and re-run the full N-body model to optimise the fit to the
observational data.  This procedure has to be done because the
location of the tidal tails differs from that of the orbit.

At outset, we do not distinguish between dark and luminous matter and
use a one-component model with a Plummer profile. Later, we also use a
more elaborate two-component model, motivated by the endpoints of
cosmological simulations.  It has a Hernquist sphere corresponding to
the luminous matter, embedded in a dark matter halo which has the
Navarro-Frenk-White form.  We investigate the effects of changing the
initial mass, dark matter content and scale-lengths of both models.

\section{The Orbit}
\label{sec:results}

\subsection{Predicted Velocities}
\label{sec:vel}

\begin{figure}
  \centering 
  \epsfxsize=8cm 
  \epsffile{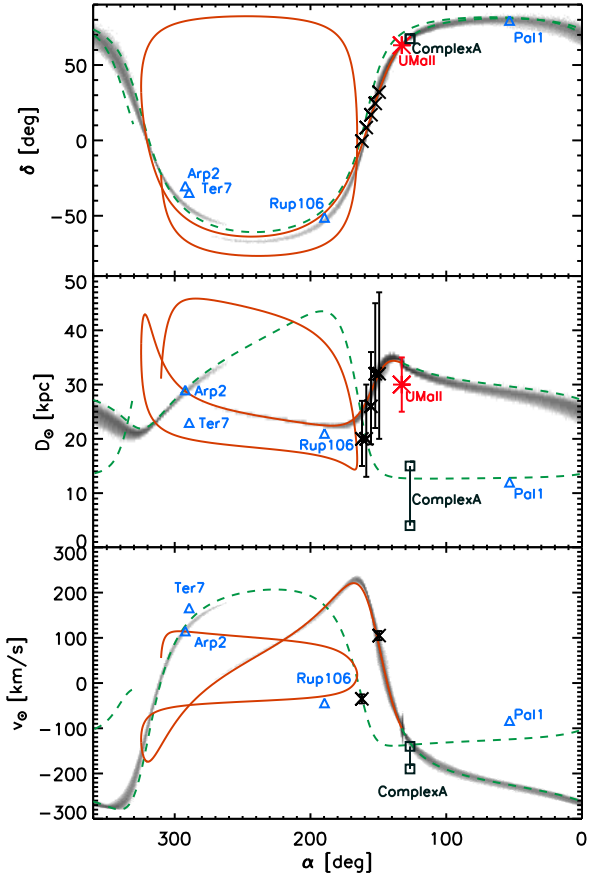}
  \caption{All-sky view of the UMa~II orbit and the best matching
  simulation. The panels show right ascension versus declination
  (upper), right ascension versus heliocentric distance (middle) and
  right ascension versus heliocentric radial velocity (lower).  The
  grey-scale contours show the logarithmic densities of the UMa~II
  tidal tails in the simulation.  The solid red (dashed green) line is
  the backward (forward) orbit shown for 1.5~Gyr. Black crosses mark
  the position of the Orphan Stream and the red star marks the
  position of UMa~II.  Blue triangles show the globular clusters,
  Arp~2, Terzan~7, Ruprecht~106 and Palomar~1. Squares show the
  position of Complex~A with its distance and velocity brackets. [The
  starting mass of the Plummer model representing UMa~II is $4 \times
  10^5$~M$_{\odot}$ and its scale-length is 80 pc]}
\label{fig:allsky}
\end{figure}

Our best matching orbit puts UMa~II at a heliocentric distance of
$34$~kpc, which agrees with the observational datum of \citet{Zu06}.
We predict the radial velocity and (heliocentric) proper motions of
UMa~II as
\begin{eqnarray}
  \label{eq:pm}
  v_\odot & = & -100 \ {\rm km\,s}^{-1},\nonumber\\ 
  \mu_{\alpha} \cos \delta & = & -0.33 \ {\rm mas}\;{\rm yr}^{-1} \\
  \mu_{\delta} & = & -0.51 \ {\rm mas}\;{\rm yr}^{-1}.\nonumber
\end{eqnarray}
The resulting orbit is shown in Fig.~\ref{fig:orbit}. It has a
perigalacticon of $\sim 18.4$~kpc and an apogalacticon of $\sim
40.6$~kpc.  This orbit not only connects UMa~II with the Orphan Stream
but also permits Complex~A and several globular clusters of the Milky
Way to be related to it.  

The tidal tails of UMa~II do not lie precisely along UMa~II's orbit.
Fig.~\ref{fig:allsky} shows grey-scale contours of the tidal debris in
the planes of right ascension versus declination, heliocentric
distance and heliocentric velocity respectively.  The positional data
on the Orphan Stream is nicely matched by the tidal tails.  The model
predicts a strong velocity gradient along the Orphan Stream with the
radial velocity varying from $200\ {\rm kms}^{-1}$ at the southern end
($\alpha \approx 170^\circ$) to $-100 \ {\rm kms}^{-1}$ at the
northern end ($\alpha \approx 130^\circ$).  The gradient in radial
velocity becomes shallower at higher declinations.

The Orphan Stream may also have been detected as a density enhancement
in star count data derived from CADIS or the Calar Alto Deep Imaging
Survey~\citep{Fu06}.  This idea receives some support from
Fig.~\ref{fig:allsky}, as their 9h field falls on the second wrap of
the backward orbit.

\subsection{Possibly Associated Objects}
\label{sec:assobj}

On the basis of intersections of their polar paths, \citet{Be06b}
speculated that there may be a connection between the Orphan Stream
and a number of anomalous, young halo globular clusters -- in
particular Palomar 1, Ruprecht 106, Arp 2 and Terzan 7.

From Figs.~\ref{fig:orbit} and ~\ref{fig:allsky}, we can assess how
Belokurov et al.'s speculations fare against the simulation.  The
position and radial velocity of Pal 1 is a good match to the forward
orbit of UMa II. In this context, it is interesting to note that
Figure 1 of \citet{Zu06} shows clumps visible in the central parts of
UMa~II. Pal 1 looks like one such clump that has already broken off
and leads UMa~II. Arp 2 is also well-matched in position and radial
velocity of the forward orbit, although it has also been claimed as a
possible Sagittarius stream member on the basis of distance,
kinematics and chemical composition~\citep[see e.g.,][]{Sb05}.  The
position of Rup 106 is a good match to the backwards orbit, but its
velocity is not (it should lie on the upper rather than the lower wrap
in Fig.~\ref{fig:allsky}).  However, bearing in mind the distance
errors to the globular clusters, Rup 106 probably cannot be discarded.
Ter 7 does seem to be ruled out -- the right panel of
Fig.~\ref{fig:orbit} and the middle panel of Fig.~\ref{fig:allsky}
show substantial mismatches between its distance and that of the
forward orbit of UMa~II.

\citet{Be06b} also pointed out the remarkable alignment between the
Orphan Stream and the Complex A association of High Velocity Clouds
(HVCs). Although we do not address the origin of Complex A in this
paper, we note that the forward orbit does pass through the location
of Complex A, and even the heliocentric velocities are reasonably
well-matched (see the lower panel of Fig.~\ref{fig:allsky}).  If the
clouds of Complex A are indeed associated with the Orphan Stream, the
simulation suggests that they lie more than a revolution ahead in orbital
phase.

\section{The Morphology of the Orphan Stream}
\label{sec:os}

\begin{figure*}
  \centering 
  \epsfxsize=8cm 
  \epsfysize=9cm 
  \epsffile{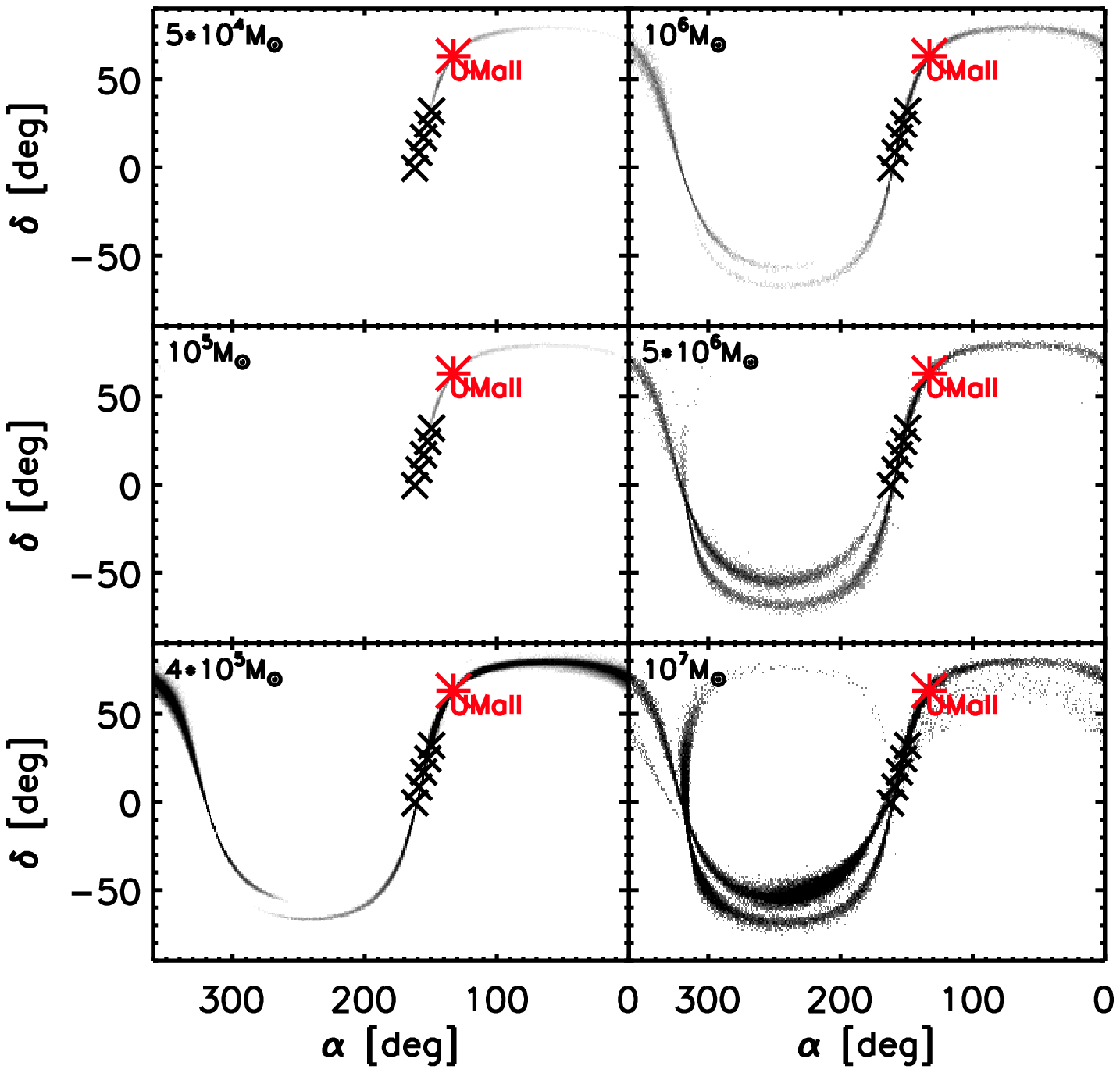}
  \epsfxsize=8cm 
  \epsfysize=9cm 
  \epsffile{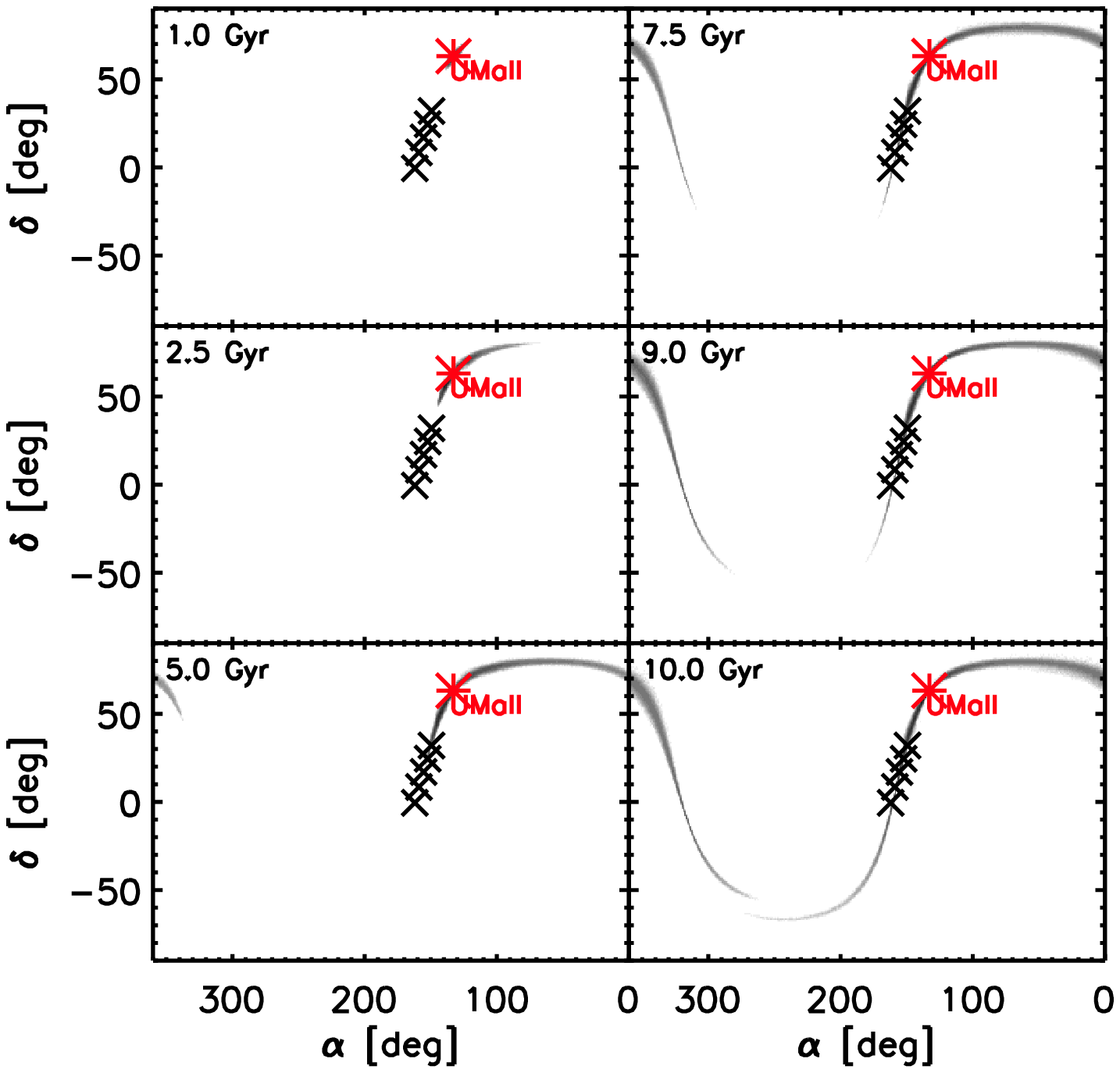}
  \caption{Left: The length of the tidal tail as a function of the
  initial mass of the object, shown in the top left of each panel. The
  red star shows the position of UMa~II, the black crosses the
  positions of the Orphan Stream. If the initial mass is of the order
  $5 \times 10^6$~M$_\odot$ or more, we should see more than one
  wrap. If the initial mass is of the order $10^5$~M$_\odot$ or less,
  the tail is not long enough to match the Orphan Stream. [The
  scale-length of the Plummer model representing the progenitor of
  UMa~II is 80 pc. The duration of the simulation is 10 Gyr.]  Right:
  The length of the tidal tails as a function of time, shown in the
  top left of each panel. The time has to be on the order of 7.5 Gyr
  or greater to ensure that the tails are long enough to match the
  Orphan Stream. [The starting mass of UMa~II is $4 \times
  10^5$~M$_{\odot}$ and its scale-length is 80~pc.] }
\label{fig:umaprogab}
\end{figure*}

\begin{figure}
  \centering
  \epsfxsize=8cm
  \epsfysize=12cm
  \epsffile{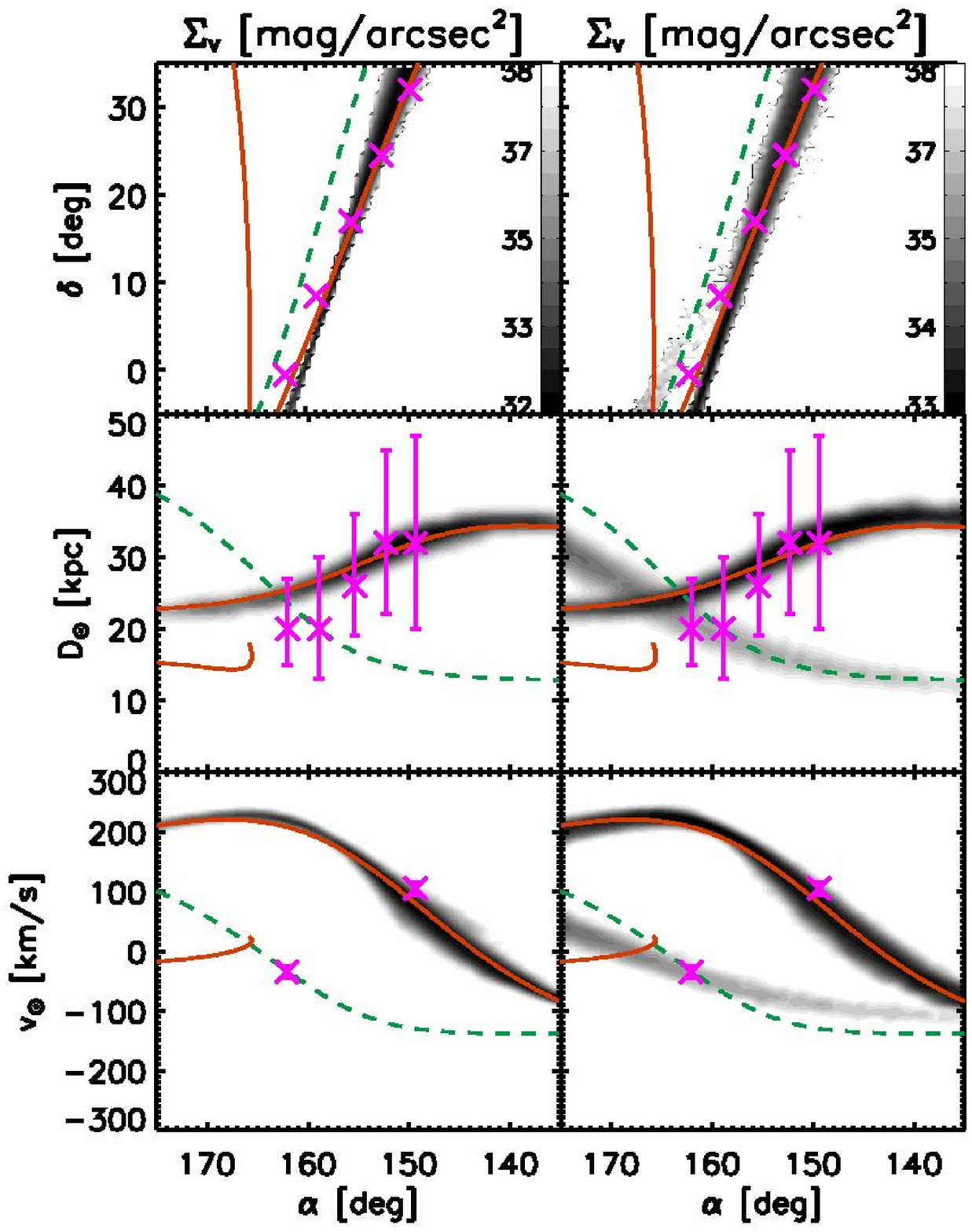}
  \caption{A close-up of the Orphan Stream in the one-component model
    (left panels) and the two-component model, showing luminous mass
    only (right panels).  Red solid (green dashed) line shows the
    backward (forward) orbit of UMa~II.  The purple crosses with
    error-bars mark the observational results from \citet{Be06b}.
    From top to bottom the panels show the surface brightness in $V$
    (mass is converted into luminosity using a mass-to-light ratio of
    $2$), the logarithmic density distribution in right
    ascension--heliocentric distance space and the logarithmic density
    distribution in right ascension--radial velocity space.  Both
    models match the positional data of the Orphan Stream.  Both
    models also fit the observational distances within the errors, but
    in the two-component model the closest two data-points are better
    matched with the wrap-around of the leading arm, which is not
    present in the one-component model.  Also, the velocity
    measurements are only matched if the leading arm is present.}
    \label{fig:oscomp}
\end{figure}

With this orbit in hand, we can deduce some constraints on the initial
mass of UMa~II.  The length of the tidal tails is controlled by the
total initial mass of the satellite (dark and luminous matter are of
course not differentiated in our one-component simulations).  If this
mass is $\lta 10^{5}$~M$_{\odot}$, the resulting tails are too short
to be consistent with the $\sim 50^\circ$ arc of the Orphan Stream
visible in SDSS. On the other hand, if the initial mass is $\gta 5
\times 10^{6}$~M$_{\odot}$, further wraps of the leading and trailing
arms should then be seen in SDSS data.  This is illustrated in the
left panels of Fig.~\ref{fig:umaprogab}, which shows the tidal tails
produced by the disruption of a sequence of UMa~IIs of different
starting masses. Similarly, sequences of the disruption of UMa~IIs for
different times, as shown in the right panels of
Fig.~\ref{fig:umaprogab}, suggest that timescales less than 7.5 Gyr
are insufficient to reproduce the present-day length of the Orphan
Stream.

Having found lower limits for the progenitor mass and the simulation
time, we now focus on two particular models.  The first is a
one-component model in which dark and luminous matter are not
distinguished. It has a Plummer distribution with a mass of $M_{\rm
pl} = 4 \times 10^{5}$~M$_{\odot}$ and a scale-length of $R_{\rm pl}=
80$~pc.  The rationale for these parameters will become clear in
Sect.~\ref{sec:remnant}, where we place further constraints on the
progenitor mass by investigating the remnant.  The second is a
two-component model with the luminous matter represented by a
Hernquist sphere with mass $M_{\rm hern} = 5 \times
10^{5}$~M$_{\odot}$ and a scale-length of $200$~pc. The is embedded in
a Navarro-Frenk-White (NFW) dark matter halo. The NFW model has the
same scale-length as the luminous matter, together with a mass within
the cut-off radius (set to be the tidal radius at perigalacticon)
which is ten times greater than the luminous mass. If the NFW mass is
made larger or the scale-length smaller, then the progenitor becomes
much harder to disrupt and does not resemble the present-day UMa~II.

Fig.~\ref{fig:oscomp} shows a close-up of the simulation data at the
position of the Orphan Stream for the one and two-component models.
There are a number of morphological features that both the simulations
reproduce successfully. First, the tidal tails of the models have a
full-width half-maximum (FWHM) of $\sim 2^\circ$.  This matches the
FWHM of the Orphan Stream as measured by \citet{Be06b}.  In both
simulations, the mass in the Orphan Stream is $\lta 10^5$ M$_\odot$,
in reasonable agreement with the stellar mass inferred from its
luminosity of $\sim 8 \times 10^4$L$_\odot$~\citep{Be06b}.  There is
just one arm visible in the one-component model, and the total mass in
the Stream is $\sim 6 \times 10^4$~M$_{\odot}$. This is a closer match
than the $\sim 3 \times 10^4$~M$_{\odot}$ in stars present in both
arms in the Orphan Stream for the two-component model.

Both models reproduce the positional data of the Orphan Stream very
well and are in good agreement with the measured distances.
Nevertheless, in the two-component model, a wrap-around of the leading
arm is present, which gives a better fit to the two low declination
data points in the middle panels of Fig.~\ref{fig:oscomp}.  Further,
the velocity data-point at the low declination (or high right
ascension) end of the Stream shown in the lower panels of
Fig.~\ref{fig:oscomp} can only be reproduced with the presence of a
wrapped around leading arm.

\begin{figure}
  \centering
  \epsfxsize=8cm
  \epsfysize=12cm
  \epsffile{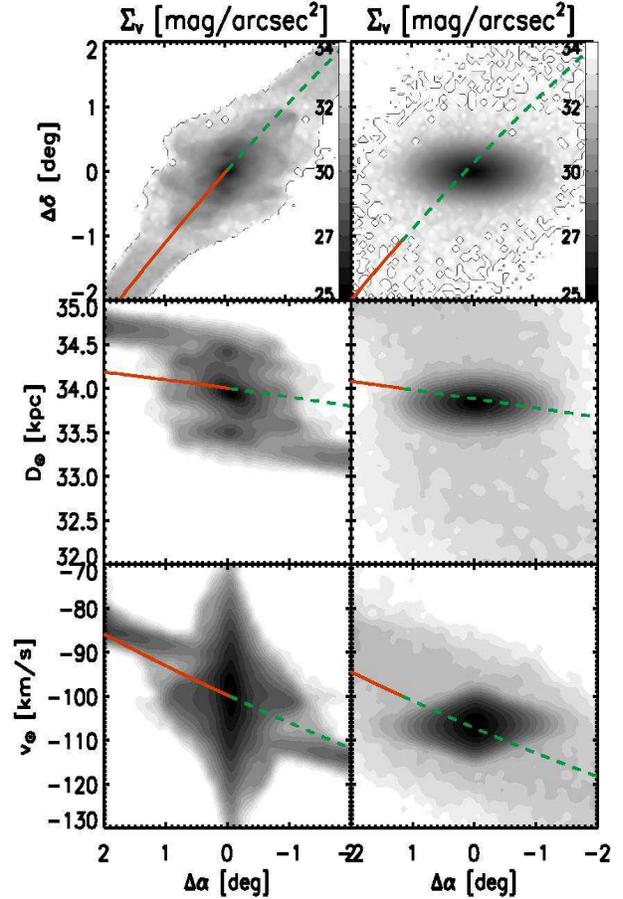}  
  \caption{A close-up of the UMa~II remnant in both models.  Left
    panels show again the one-component model, while right panels show
    the two-component model.  The top row shows the surface brightness
    of all particles in the one-component model and of the luminous
    matter only in the two-component model.  The second row gives the
    logarithmic density distribution in right ascension--distance
    space while the third row shows the distributions in right
    ascension--radial velocity. The coloured lines are as in
    Fig.~\ref{fig:oscomp}.}  \label{fig:umacomp}
\end{figure}
\begin{figure}
  \centering
  \epsfxsize=5cm
  \epsfysize=5cm
  \epsffile{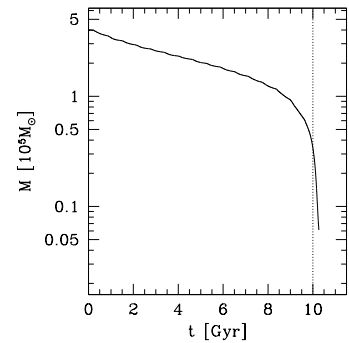}
  \caption{The bound mass of UMa~II is plotted against time for the
    one-component model.  During most of its lifetime, the mass
    decreases only slightly with each perigalacticon passage. However,
    the last disc shock leads to the final disruption of the object
    and the bound mass drops quickly to zero.  At this particular
    instant, the stars of the object become unbound but have not yet
    dispersed from the location of the object.  At later epochs, the
    bound mass is zero and the stars disperse into the tails.}
    \label{fig:mass}
\end{figure}
\begin{figure}
  \centering 
  \epsfxsize=7cm \epsfxsize=7cm
  \epsffile{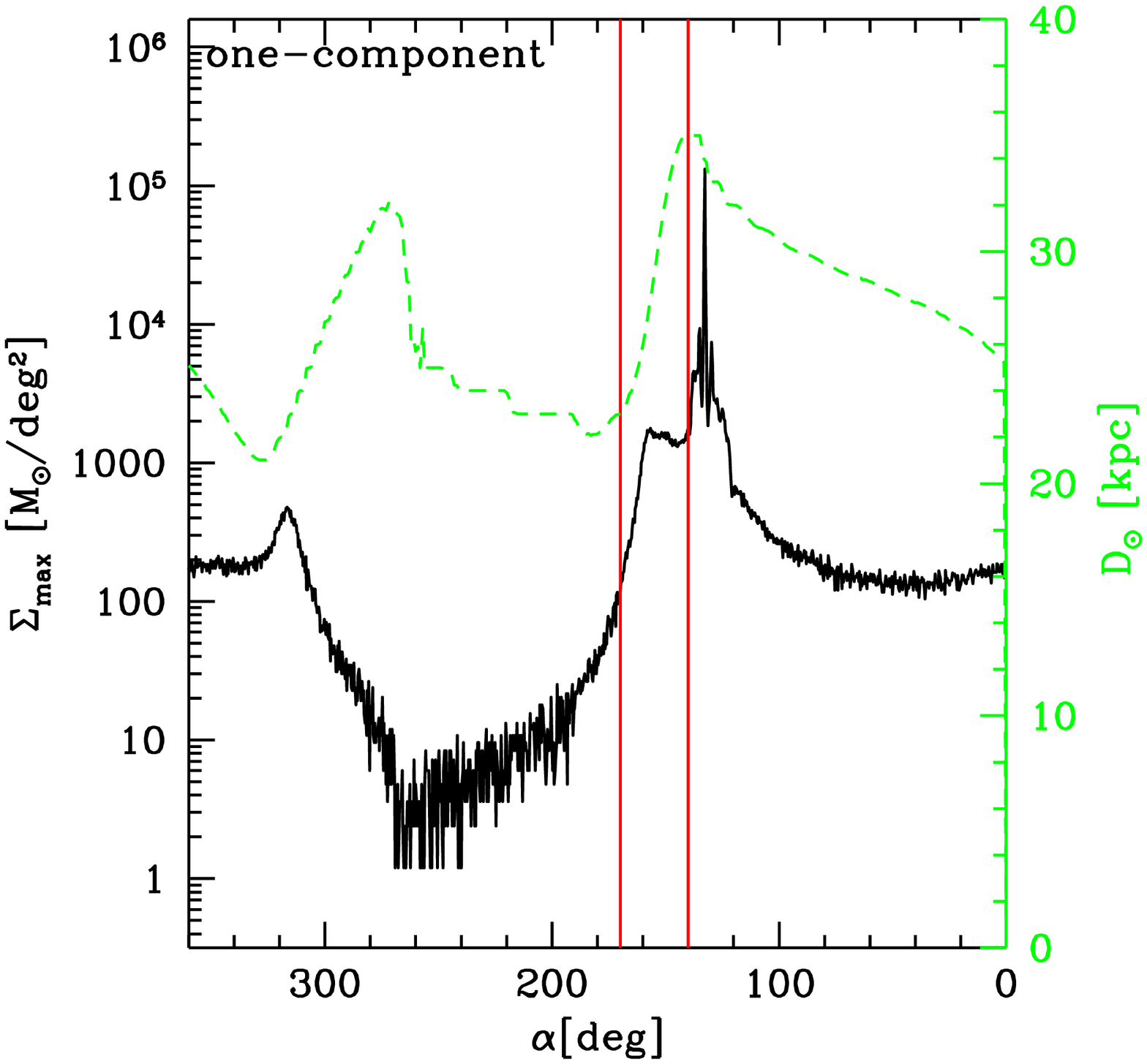} 
  \epsfxsize=7cm \epsfxsize=7cm
  \epsffile{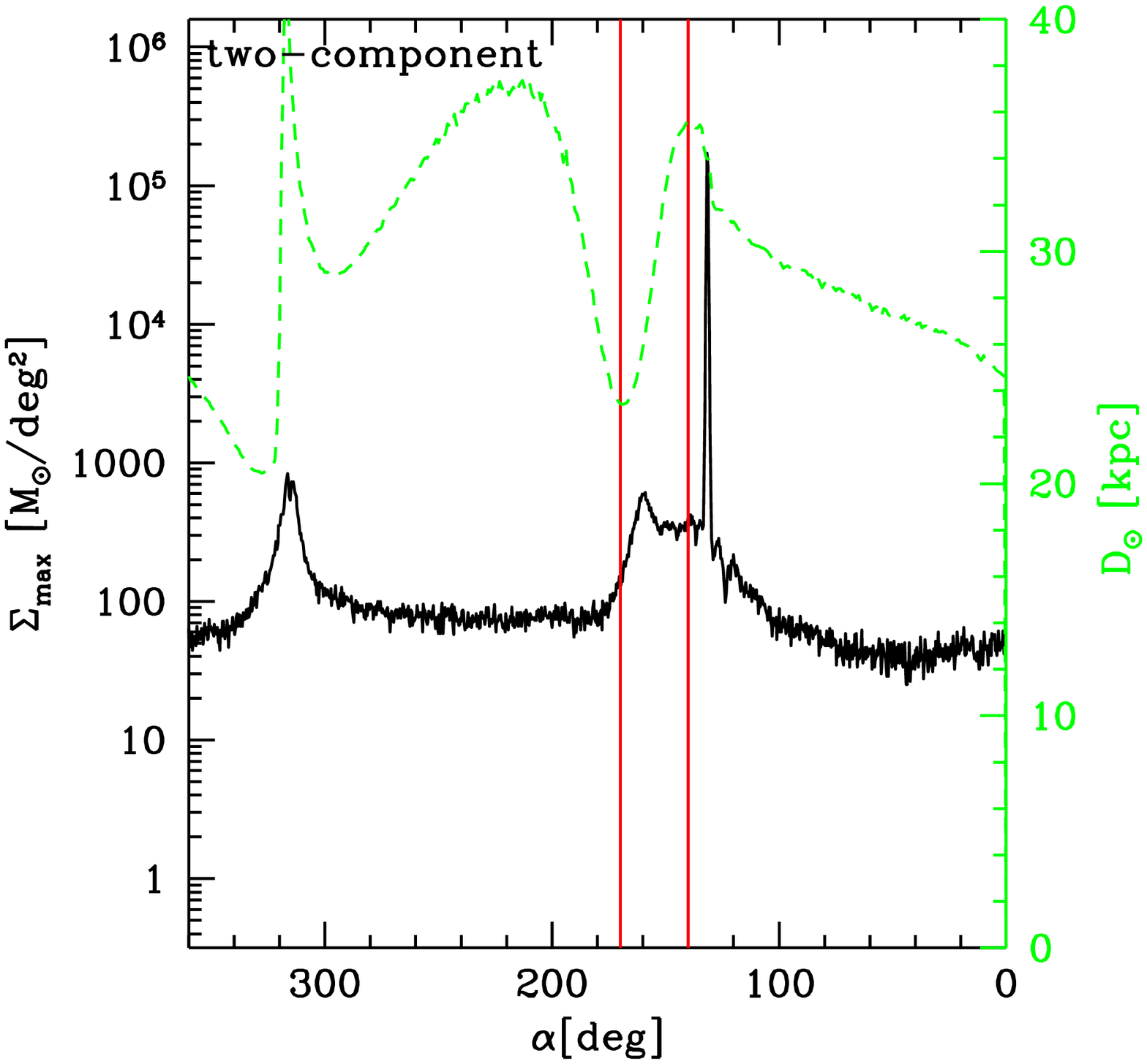}
\caption{The peak surface density $\Sigma_{\rm max}$ (black) and the
mean heliocentric distance $D_\odot$ (green) of the Orphan Stream are
plotted as a function of right ascension for the one-component
(two-component) models in the upper (lower) panels. The vertical red
lines mark the range of right ascension over which the Orphan Stream
is detected in SDSS data. UMa~II corresponds to the sharp density peak
at right ascension $\alpha = 132.8^{\circ}$.  The fading of the Orphan
Stream just before it approaches UMa~II is caused by the decreasing
peak surface density and the increasing mean distance.}
\label{fig:densOS}
\end{figure}
\begin{figure}
  \centering 
  \epsfxsize=7cm 
  \epsfxsize=7cm 
  \epsffile{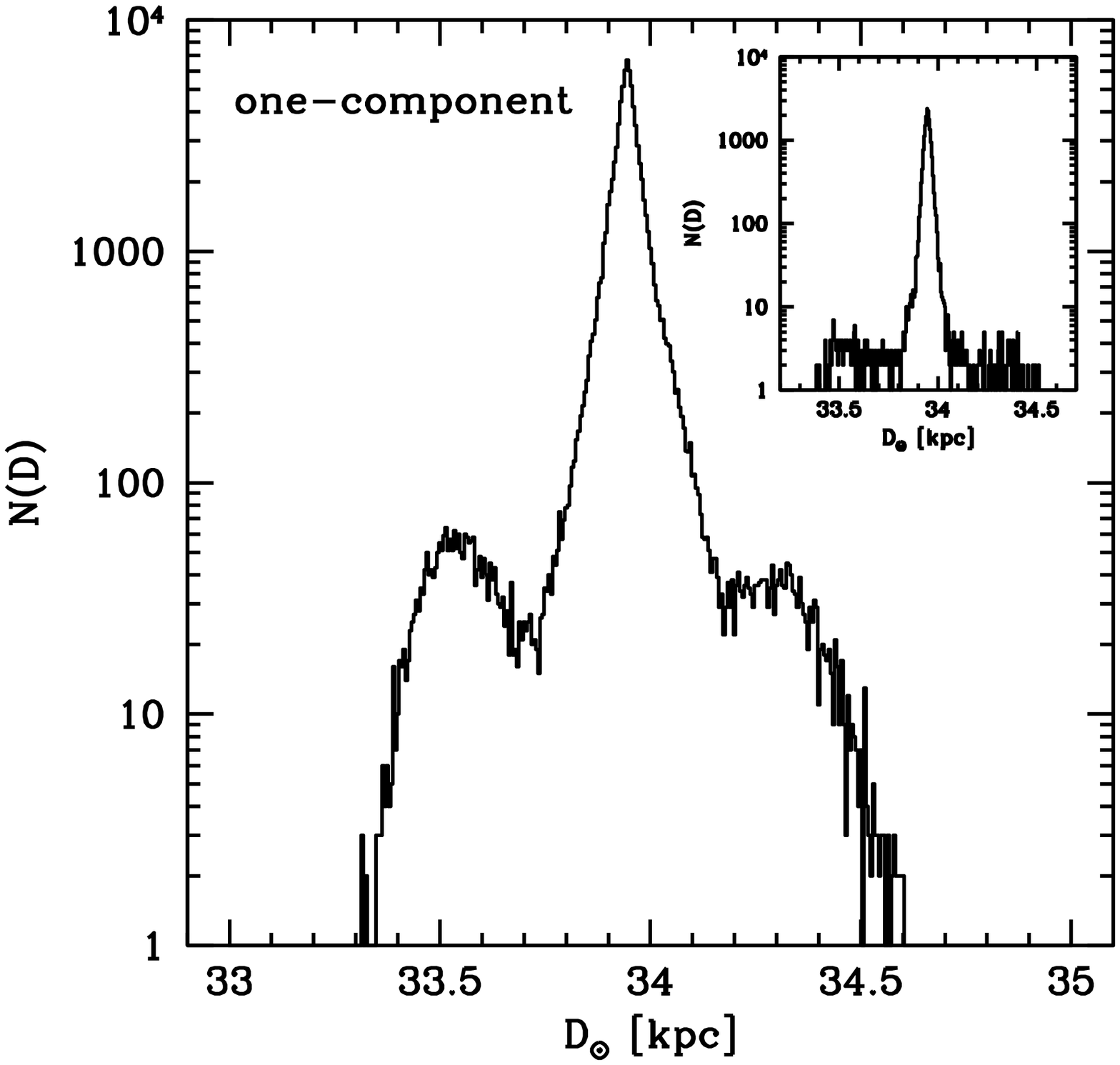} 
  \epsfxsize=7cm 
  \epsfxsize=7cm 
  \epsffile{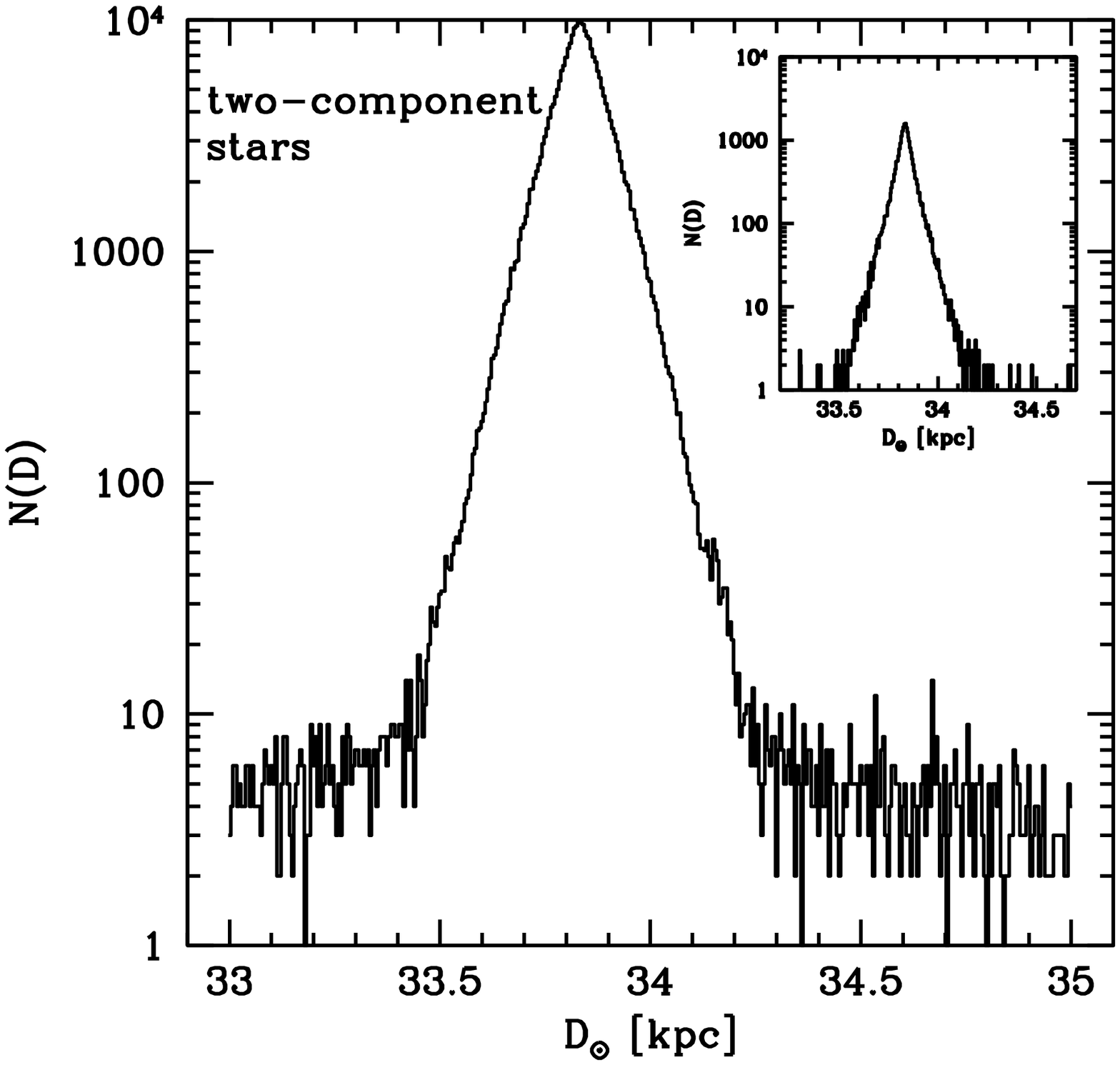} 
\caption{Histogram of heliocentric distances of stars in a $0.4^\circ
\times 0.4^\circ$ field centred on UMa~II, approximately the same size
as the panels in \citet{Zu06}. The inset shows the histogram of
distances but now confined to the very central parts of UMa~II
($0.1^\circ \times 0.1^\circ$ field).  The data are taken from the
one-component (two-component) model in the upper (lower) panels.}
\label{fig:densUM}
\end{figure}

\section{The Morphology of UMa~II}
\label{sec:remnant}

We can sharpen the constraints on the initial mass by requiring that
the simulations also reproduce the disrupted nature of UMa~II itself.
Fig.~\ref{fig:umacomp} shows the results of the disruption of the
one-component model (left panels) and the two-component model (right
panels). The three rows show the surface brightness, the logarithmic
density distribution in right ascension-heliocentric distance and in
right ascension-heliocentric velocity space, respectively.

For one-component models, we find that satellites with an initial mass
$\gta 10^{6}$~M$_{\odot}$ do not become sufficiently dissolved to
resemble the present-day UMa~II.  Below this, there is a trade-off
between starting mass and scale-length.  For example, a Plummer sphere
with mass $M_{\rm pl} = 5 \times 10^{5}$~M$_{\odot}$ and scale-length
$R_{\rm pl} = 100$~pc gets completely dissolved without a remnant,
whilst one with $R_{\rm pl} = 85$~pc gives a remnant which is too
massive by two orders of magnitude. Reducing the mass to $M_{\rm pl} =
4 \times 10^{5}$~M$_{\odot}$ and using $R_{\rm pl} = 80$~pc results in
a remnant with similar mass and aspect to UMa~II.  Fig.~\ref{fig:mass}
shows the evolution of the bound mass of our one-component model.  If
the final mass of the remnant is $\sim 6 \times 10^3$~M$_{\odot}$,
then an object with an initial mass of $\sim 10^{5}$~M$_{\odot}$ must
be in its final stage of dissolution.  A robust result is that the
initial distribution of the satellite cannot be very concentrated,
otherwise there is insufficient mass loss to produce the Orphan
Stream.

For two-component models, it is a challenge to reproduce the dissolved
nature of the present-day UMa~II.  As our illustrative example, we use
a Hernquist sphere of $5 \times 10^{5}$~M$_{\odot}$ and scale-length
of $200$~pc, embedded in a Navarro-Frenk-White halo with the same
scale-length and with a mass of $5 \times 10^{6}$~M$_\odot$ within the
tidal radius of $400$~pc.  These parameters are set given the
constraint that the present-day mass in the Orphan Stream is $\sim
10^5$~M$_\odot$.  If the mass-to-light ratio of the progenitor is
$\sim 10$, this fixes the halo mass, whilst the scale-lengths must be
in excess of $200$~pc to allow for enough luminous matter to be
stripped off and found in the Orphan Stream.  Even so, at the endpoint
of the simulation, the remnant has $\sim 10^5$~M$_{\odot}$ in stars.
This is too large by two orders of magnitude!

For both the one and two component models, the UMa~II remnant in
Fig.~\ref{fig:umacomp} shows a prominent elongation -- not along its
orbit -- but along lines of constant declination.  The same elongation
is found in the deeper, follow-up observations with the Subaru
telescope reported by \citet{Zu06}.  Comparing the size of the
observed UMa~II of about one degree along constant declination with
our models, we conclude that our one-component model fits the
extension of the real object much better than the two-component model.

\begin{figure*}
  \centering \epsfxsize=13cm \epsfysize=13cm \epsffile{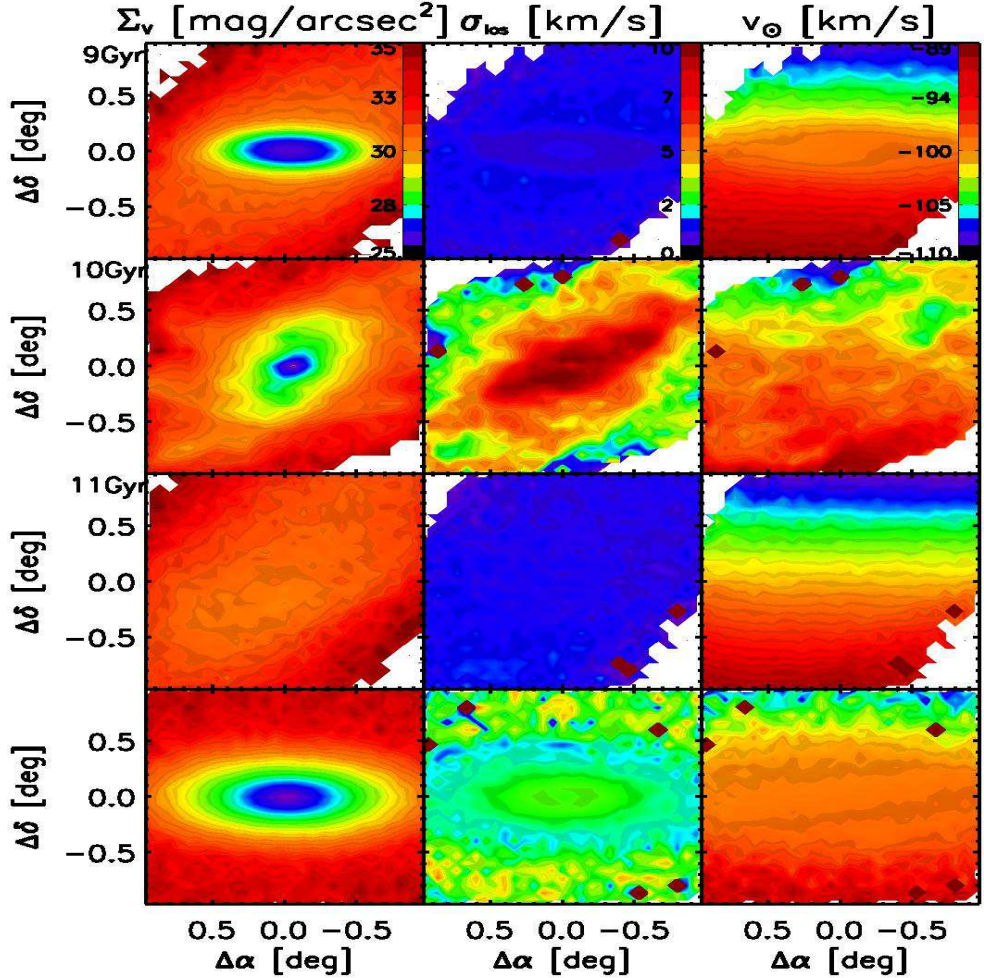}
  \caption{This shows contours of the V band surface brightness, the
  velocity dispersion and the mean radial velocity of the remnant in
  our simulations. The first three rows refer to the one-component
  model after 9~Gyr, 10~Gyr and 11~Gyr. This is a sequence from bound
  through disrupting to almost completely dissolved object. The final
  row shows the endpoint of the two-component model after 10~Gyr for
  comparison. (The key to the colour code is given on the right-side
  of the top panels).}  \label{fig:compare}
\end{figure*}

\begin{table*}
  \centering
  \begin{tabular}{lccccccc} \hline
    Galactic Model & $q$ & $D_{\odot}$ & $v_{\odot}$ & $\mu_{\alpha}$ &
    $\mu_{\delta}$ & $R_{\rm p}$ & $R_{\rm a}$ \\
     & & [kpc] & [km\,s$^{-1}$] & [mas\,yr$^{-1}$] & [mas\,yr$^{-1}$]
     &  [kpc] & [kpc] \\ \hline \hline
     Miyamoto-Nagai disc, Logarithmic halo & 0.90 & 30 & -100 & -0.50 & -0.50 & 18.9 & 34.5 \\
     Miyamoto-Nagai disc, Logarithmic halo & 1.00 & 30 & -115 & -0.40 & -0.50 & 18.8 & 37.2 \\
     Miyamoto-Nagai disc, Logarithmic halo & 1.11 & 30 & -105 & -0.35 & -0.50 & 17.8 & 36.7 \\\hline
     Dehnen \& Binney models & 1.00 & 33 & -125 & -0.25 & -0.65 & 14.2 & 40.8 \\ \hline
  \end{tabular}
  \caption{Parameters for best fit test-particle orbits for different
    choices of potential.  The first column gives the Galactic model.
    For the logarithmic halos, $q$ is the flattening of the
    equipotentials [see eq.~(\ref{eq:halopot})], whereas for the
    Dehnen \& Binney (1998) models, $q$ is the flattening of the
    isodensity contours [see eq~(\ref{eq:bin})]. The columns give the
    best-fit initial conditions of UMa~II today so as to join up with
    the Orphan Streaml; the heliocentric distance, heliocentric radial
    velocity and proper motion in $\alpha$ and $\delta$.  The last two
    columns show the peri- and apogalacticon distances of the orbit.
    Note that the values for the $q=1$ logarithmic halo case differ
    slightly from the values in the main paper because only the
    test-particle orbit was fit to the data in this Table.}
  \label{tab:extra}
\end{table*}

Another advantage of the one-component model is that there is some
substructure in the UMa~II remnant, as is visible in the upper left
panel of Fig.~\ref{fig:umacomp}. In the simulation, this is caused by
tidal shocking of the remnant at the last few disk passages.  The
substructure is qualitatively similar to internal clumpiness of the
UMa~II dSph seen by \citet{Zu06}.  However, to recover the details of
this feature may well require a more elaborate starting model than a
simple Plummer sphere.  The two-component model shows no substructure
at the end of the simulation.

Both simulations not only match all the available positional data, but
-- more strikingly -- they also explain why the tails around UMa~II
are faint and undetectable with SDSS.  At the positions of the Orphan
Stream, projection effects enhance the visibility of the
well-collimated stream, which lies almost along the line of sight.  By
contrast, at UMa~II the orbit is almost transverse to the line of
sight and there is no enhancement from projection effects.  This
provides a natural explanation as to why an extension of the Orphan
Stream is not visible all the way up to the position of UMa~II in SDSS
data.  This phenomenon is also illustrated in Fig.~\ref{fig:densOS}
which shows the peak density in the simulated Orphan Stream dropping,
and the mean heliocentric distance increasing, as UMa~II is
approached.  

Fig.~\ref{fig:densUM} shows the distribution of heliocentric distances
of stars in a $0.4^\circ \times 0.4^\circ$ field centred on UMa~II
remnant, together with an inset that records the same information but
now confined to the very innermost $0.1^\circ \times 0.1^\circ$ field.
Although the innermost parts are quite confined, the entire object has
a significant depth along the line of sight of $\sim 1$ kpc,
particularly in the one-component model. This is of the right order
of magnitude to cause the broadening of features of the
colour-magnitude diagram discerned by \citet{Zu06}.

\section{Velocity dispersions}
\label{sec:disp}

We have shown that the initial mass (stars and possible dark matter)
must exceed $10^{5}$~M$_{\odot}$ to account for the length of the
tidal tails and the known stellar mass in the Orphan Stream.  But any
object with a total mass $> 10^{6}$~M$_{\odot}$ typically leads to a
present-day UMa~II which is still strongly bound and has a luminous
matter contribution at least an order of magnitude larger than the
observed $6 \times 10^3$~M$_{\odot}$. 

One solution to this dilemma is to postulate that we are observing
UMa~II at a time close to its disintegration.  The last disc passage
led to the almost complete disruption of the remnant object.  The
stars are now rapidly becoming unbound.  They have not yet dispersed
along the orbit into the tidal tails and we still see them in the
innermost parts of UMa~II in a very confined area (see the inset of
Fig.~\ref{fig:densOS}). If so, then the interpretation of kinematic
data may need special care.

The first three rows of Fig.~\ref{fig:compare} all show one-component
models.  We used exactly the same set-up, but started the simulation
at 9~Gyr, 10~Gyr or 11~Gyr ago on the same orbit to make sure all
models are now seen at the same position on the sky (to exclude
projection effects). The bound object in the first row has a small
velocity dispersion, which is even lower than that in the surrounding
tails.  But the mean line of sight velocity is constant throughout the
bound object and a gradient is only visible in the tails.  This
changes dramatically in the disrupting model shown in the middle
panel.  We still see an object with a similar total surface
brightness, but it already shows sub-structure on small scales.  The
velocity dispersion is inflated by a factor of ten, but the dissolved
nature of the remnant is already visible in the mean radial velocity.
There is a strong gradient throughout the object, even though the mean
velocity shows some flocculent structure.  In the third row, in which
the process of disruption is almost complete, the dissolved object has
a low surface density, which will decrease further in the future until
it matches that of the tails.  The velocity dispersion is again low,
at much the same value as that of the tails.  Looking at the mean
radial velocity, it is hard to distinguish what remains of the object
from the tails.  For comparison, the final row of
Fig.~\ref{fig:compare} shows the same quantities for the two-component
model. The final bound object has smooth surface brightness contours,
a high velocity dispersion because of the dark matter content and no
gradient in the mean radial velocity.

Follow-up high precision kinematic observations of this new dwarf
galaxy could reveal a high velocity dispersion, irrespective of the
dark matter content. However, the existence of a gradient in the mean
radial velocity provides a clear-cut distinction between a disrupting
object and a bound, dark matter dominated object.

\section{The Galactic Potential}
\label{sec:extra}

Hitherto, our Galactic model is built from three fairly simple
analytic components that could have some deficiencies. Although we are
using the same standard model as many previous
investigators~\citep[e.g.,][]{He04,Jo05}, it is prudent to examine the
robustness of our results to changes in the underlying Galactic
potential. Whilst re-running all the N-body simulations would be
time-consuming, it is straightforward to carry out the initial
test-particle calculations described in Section 2 for different
potentials.

For example, we can vary the flattening of the halo.  If we change the
halo shape from spherical to moderately prolate or oblate, we still
are able to fit all the data by slightly altering the starting
velocities of UMa~II. This is illustrated in Table~\ref{tab:extra},
which gives the velocities, and the pericentric and apocentric
distances for test particle calculations. Note that $q$ in the
logarithmic halo refers to the flattening of the equipotentials -- the
flattening in the density contours is typically two or three times
greater~\citep[see e.g.,][]{Ev93}.  For moderate changes, a suitable
orbit can always be found that joins up UMa~II with the Orphan Stream,
but UMa~II's predicted velocity and proper motions are then somewhat
different. Only if we use strongly prolate or oblate models does the
orbit of UMa~II change so dramatically that we are not able to fit all
the data on the Orphan Stream at once. However, the recent study of
the multiple wraps of the Sagittarius' stream by \citet{Fe06} provides
strong evidence that only spherical or close to spherical halo shapes
are possible for the Milky Way.

As a further check, we change the type of the Galactic potential and
use a Dehnen \& Binney (1998) model.  These potentials consist of
three exponential discs (thin, thick and gaseous). The halo and the
bulge are represented by two spheroidal distributions
\begin{eqnarray}
  \label{eq:bin}
  \rho_{\rm S}(R,z) & = & \rho_{0} \left( \frac{m}{r_{0}}
  \right)^{-\gamma} \left( 1 + \frac{m}{r_{0}} \right)^{\gamma -
    \beta} \exp \left( - \frac{m^{2}} {r_{\rm t}^{2}} \right).
\end{eqnarray}
Here $m^2 = R^2 + z^2 q^{-2}$ and $q$ is the axis ratio in the
density, whilst the remaining parameters are chosen as
in~\citet{Fe06}. Table~\ref{tab:extra} shows how the intitial
conditions or the test particle orbit change for the spherical case
($q\!=\!1$) for comparison. Again, an orbit matching UMa~II to the Orphan
Stream can be found, and the changes in the initial conditions in
Table~\ref{tab:extra} give an indication of the likely uncertainties
in our predictions caused by changes in the Galactic potential.

\section{Conclusions}
\label{sec:conclus}

We have carried out N-body simulations to model the evolution and
disruption of the recently discovered dwarf galaxy UMa~II. The
simulations reproduce the available observational data on the Orphan
Stream within their error margins.  We conclude that UMa~II is a
likely progenitor of the Orphan Stream.  We predict the radial
velocity of UMa~II as $-100 \ {\rm km\,s}^{-1}$.  We also predict a
strong velocity gradient along the Orphan Stream with the radial
velocity varying from $\sim 200\ {\rm kms}^{-1}$ at the southern end
to $\sim -100 \ {\rm kms}^{-1}$ at the northern end.

From the length of the tails and the mass found in the Orphan Stream,
we deduce that the initial mass of UMa~II is in excess of
$10^{5}$~M$_{\odot}$.  But, an object more massive than
$10^{6}$~M$_{\odot}$ cannot be dissolved to produce the present day
UMa~II, at least on the orbit derived from the observations.
Therefore, the initial mass of UMa~II has to be of the order a few
$10^{5}$~M$_{\odot}$. To reduce UMa~II's mass through tidal effects to
its present value, the distribution of stars and dark matter has to be
extended. We carried out a suite of simulations of the disruption of
UMa~II with one-component models, which have little dark matter beyond
that associated with the stellar populations, and two-component models
with a mass-to-light ratio of $\sim 10$. There are strengths and
weaknesses of both sets of simulations. Both reproduce the positions
and distances of the Orphan Stream, but the two-component models are
in better agreement with the admittedly uncertain kinematic data
derived by \citet{Be06b}. However, the one-component models can
provide a much better match to the disrupted nature of UMa~II today.
The velocity dispersion is not a clean test between these two
possibilities, as we have shown that objects undergoing disruption can
have an anomalously high velocity dispersion. However, a clear-cut
test is provided by the mean radial velocity, which should show no
gradient for dark matter dominated models, but an obvious gradient for
disrupting models.

The orbit that we have derived supports the idea of \citet{Be06b} that
some of the anomalous, young halo globular clusters (particularly
Pal~1, Arp~2 and possibly Rup~105) may be associated with the Orphan
Stream.  Intriguingly, the position and velocity of Complex~A can also
be matched, but only if it lies a revolution ahead in orbital
phase. The association of these objects however makes most sense in
the picture in which UMa~II, the young halo globular clusters and
Complex~A are all fragments of a much larger object like a tidal dwarf
galaxy.

\section*{Acknowledgments}

MF, VB, DBZ, MIW and DMB gratefully acknowledge financial support
through PPARC.  Funding for the SDSS and SDSS-II has been provided by
the Alfred P.  Sloan Foundation, the Participating Institutions, the
National Science Foundation, the U.S. Department of Energy, the
National Aeronautics and Space Administration, the Japanese
Monbukagakusho, the Max Planck Society, and the Higher Education
Funding Council for England. The SDSS Web Site is
http://www.sdss.org/.

The SDSS is managed by the Astrophysical Research Consortium for the
Participating Institutions. The Participating Institutions are the
American Museum of Natural History, Astrophysical Institute Potsdam,
University of Basel, Cambridge University, Case Western Reserve
University, University of Chicago, Drexel University, Fermilab, the
Institute for Advanced Study, the Japan Participation Group, Johns
Hopkins University, the Joint Institute for Nuclear Astrophysics, the
Kavli Institute for Particle Astrophysics and Cosmology, the Korean
Scientist Group, the Chinese Academy of Sciences (LAMOST), Los Alamos
National Laboratory, the Max-Planck-Institute for Astronomy (MPIA),
the Max-Planck-Institute for Astrophysics (MPA), New Mexico State
University, Ohio State University, University of Pittsburgh,
University of Portsmouth, Princeton University, the United States
Naval Observatory, and the University of Washington. MF, VB, DZ and MW
thankfully acknowledge financial support through PPARC.

\label{lastpage}


\begin{thebibliography}{99}

\bibitem[Belokurov et al.(2006a)]{Be06a} Belokurov V., et al., 2006a,
  ApJ, 642, L137 

\bibitem[Belokurov et al.(2006b)]{Be06b} Belokurov V., et al., 2006b,
  ApJ, in press (astro-ph/0605705)

\bibitem[Dehnen \& Binney(1998)]{de98}
  Dehnen W., Binney J., 1998, MNRAS, 294, 429

\bibitem[Evans(1993)]{Ev93} Evans, N.~W.\ 1993, MNRAS, 260, 
191 

\bibitem[Fellhauer et al.(2000)]{Fe00} Fellhauer M., Kroupa P.,
  Baumgardt H., Bien R., Boily C.M., Spurzem R., Wassmer N., 2000,
  NewA, 5, 305 

\bibitem[Fellhauer et al.(2006)]{Fe06} Fellhauer M., et al., 2006,
  ApJ, 651, 167

\bibitem[Fuchs et al.(2006)]{Fu06} Fuchs B., Phleps S., Meisenheimer
K., 2006, A\&A, 457, 541

\bibitem[Grillmair(2006)]{Gr06} Grillmair, C., 2006, ApJ, 645, L37

\bibitem[Helmi(2004)]{He04} Helmi, A.\ 2004, MNRAS, 351, 
643 

\bibitem[Johnston et al.(2005)]{Jo05} Johnston, K.~V., Law, D.~R., \&
Majewski, S.~R.\ 2005, ApJ, 619, 800

\bibitem[Kroupa(1997)]{Kr97} Kroupa, P., 1997, New Astronomy, 2, 139

\bibitem[Sbordone et al.(2005)]{Sb05} Sbordone L., Bonifacio P.,
  Marconi G., Buonanno R., Zaggia S., 2005, A\&A, 437, 905

\bibitem[Wakker(2001)]{Wa01}
Wakker, B., 2001, ApJS, 136, 463

\bibitem[Wakker et al.(1996)]{Wa96} Wakker B., Howk C., Schwarz, U.,
van Woerden, H., Beers, T., Wilhelm, R., Kalberla, P., Danly, L.,
1996, ApJ, 473, 834

\bibitem[York et al.(2000)]{Yo00} York D.G., et al., 2000, AJ, 120,
  1579

\bibitem[Zucker et al.(2006)]{Zu06} Zucker D.B., et al., 2006,
  ApJ, 650, L41

\end{thebibliography}
\end{document}